\documentclass{article}

\usepackage{arxiv}

\usepackage[utf8]{inputenc} 
\usepackage[T1]{fontenc}    
\usepackage{hyperref}       
\usepackage{url}            
\usepackage{booktabs}       
\usepackage{amsfonts}       
\usepackage{nicefrac}       
\usepackage{microtype}      
\usepackage{lipsum}
\usepackage{graphicx}

\title{Behavioral Petri Net Mining and Automated Analysis for Human-Computer Interaction Recommendations in Multi-Application Environments}

\author{
  Julian Theis \\
  Department of Mechanical and Industrial Engineering\\
  University of Illinois at Chicago\\
  Chicago, IL 60607 \\
  \texttt{jtheis3@uic.edu} \\
   \And
 Houshang Darabi \\
  Department of Mechanical and Industrial Engineering\\
  University of Illinois at Chicago\\
  Chicago, IL 60607 \\
  \texttt{hdarabi@uic.edu} \\
}

\begin{document}
\maketitle

\begin{abstract}
Process Mining is a famous technique which is frequently applied to Software Development Processes, while being neglected in Human-Computer Interaction (HCI) recommendation applications. Organizations usually train employees to interact with required IT systems. Often, employees, or users in general, develop their own strategies for solving repetitive tasks and processes. However, organizations find it hard to detect whether employees interact efficiently with IT systems or not. Hence, we have developed a method which detects inefficient behavior assuming that at least one optimal HCI strategy is known. This method provides recommendations to gradually adapt users' behavior towards the optimal way of interaction considering satisfaction of users. Based on users' behavior logs tracked by a Java application suitable for multi-application and multi-instance environments, we demonstrate  the applicability for a specific task in a common Windows environment utilizing realistic simulated behaviors of users. 
\end{abstract}

\keywords{Multi-Application Environments \and Software Process Mining \and User Behavior Optimization \and Human-Computer Interaction Recommendation \and Behavioral Petri Nets}

\section{Introduction}\label{sec:introduction}
Today's businesses are highly dependent on software which gets more and more complex, especially when applied in industrial environments \cite{TowardsModernInclusiveFactories}. Moreover, IT expertise is perceived as self-evident in many professions. In industries like Communication, Healthcare, Education, and Government, employees are often confronted with similar and repetitive software processes. Usually, the UIs of such software are developed to be used by a broad audience. Hence, there is a chance that the users in a given organization select to use a UI in a nonoptimal way. Therefore, organizations need to train their users to use the UI in the best possible way. How users deal with a corresponding UI can be seen via UI interaction event log analysis. Standard Human-Computer Interactions (HCI) such as mouse clicks, mouse movements, and keystrokes can be tracked and the called application function information can be stored in log files. These logs, in turn, can be analyzed by Process Mining techniques to detect patterns. 

Process Mining is the extraction of process models from event logs \cite{VDAalst}. Analyzing logs originating from software environments aiming on optimizing software design, running behavior, and user-software interaction is also referred to as \textit{Software Process Mining} \cite{PMtoSW, LiuTwoLayeredFramework}. Software Process Mining plays an essential role in UI design. In many companies, users do not use the available UIs efficiently. In such cases, there is a need to train the users to use the company's UIs in a more efficient, and perhaps optimal, way. However, organizations do not always have resources to train each employee separately. A more cost-effective solution would be to train employees using individual recommendations provided by a software.

In this paper, we are proposing a method to detect inefficient user's behavior interactions in a multi-instance and multi-application environment assuming that at least one optimal way of interaction is known. Our method provides recommendations to gradually train employees or users. Potential applications can be found in areas in which interaction speeds are crucial - such as Electronic Health Record (EHR) systems. Doctors and nurses are often confronted with time issues when using EHR systems. Using such systems in a faster way saves time which can be spent on patients and/or reduce stress and workload. Similarly, our approach can be applied to train receptionist at hotels, helpdesk employees in financial institutions, or to optimize spreadsheet operations of users in virtually any industry. 

To the best of our knowledge, we are the first ones applying Process Mining to provide HCI recommendations in such environments. Moreover, we provide recommendations ranked by average time saving such that a user gradually reaches to an optimal way of interaction. 

After providing a brief introduction, we present the background and related work in Section \ref{sec:relatedwork}. We provide relevant preliminaries and definitions in Section \ref{sec:preliminaries}. The experimental and simulation environment for data acquisition purposes and the actual recommendation engine are introduced in Section \ref{sec:approach}. We discuss the obtained results in Section \ref{sec:results} and conclude the paper in Section \ref{sec:conclusion}.

\section{Background and Related Work}\label{sec:relatedwork}
When working with UIs, one should always keep in mind the overall objectives. Naively, one might think the design of an application should be modern, generate interest, and be easy to use. Though these assumptions are usually true, UIs have a much larger impact since almost all businesses are highly dependent on software tools for their success \cite{ERP,Park2010}. Intelligently designed UIs reduce the number of errors and their related cost \cite{Shneiderman:2009:DUI:1593001}. They also reduce the task interaction time, resulting in an overall higher productivity \cite{TaskBasedUIDesign}. 
In industrial applications, these task interaction times are production times due to reduced iterations and rework. Moreover, an efficient UI does not only affect the software's usability, but also leads to a smooth completion of any task at hand thereby making the interaction satisfactory as per the requirements of the user. In turn, a complicated or impractical UI can be an existing problem within an organization. Exemplary, an inefficient UI can slow down operations and eventually cause stress for employees \cite{CALISIR2004505}. It is also important to note that employees might not always complain about UI inefficiencies making the detection even harder. Therefore, companies are highly interested in conducting application log analysis to unveil frequent users' tasks and to understand users' behavior \cite{Dev:2017:IFU:3025171.3025184}.
In any UI and HCI problem, one always wants to detect an optimal solution for a specific problem, i.e. an optimal way of interaction where the problem is the detection of an optimal user's behavior in a software environment. In this case an optimal user's behavior can be defined as one that has the maximum \textit{Effectiveness}, \textit{Efficiency}, and \textit{Satisfaction} \cite{Cronholm:2009:UUG:1738826.1738864, Nielsen:1994:UE:2821575, Jacko:2012:HIH:2378709}. 

Formal methods, such as Process Mining techniques, have been used for usability analysis and user assistance for a long time to achieve effective, efficient and satisfying UIs. However, they have not been applied to provide automatic HCI recommendations. For example, Hartman et al. proposed a method called AUGUR to assist users in navigating and entering data in form applications \cite{AUGUR}. However, this method applies only to web-based applications and does not consider the interaction between multiple applications.

Palanque et al. proposed a formal method consisting of a set of different techniques to facilitate the evaluation of usability \cite{Palanque}. Their approach is to model applications using high-level PNs and to evaluate the application using observed user's logs. From the user's logs and replay on the application PN, one can observe failures of the task as well as usability issues. This approach can detect usability issues. However, it cannot be used to train the users. It cannot be used for user's behavior evaluation either. Moreover, this method requires a modeling of the application in a non-automated way.

Thimbleby et al. have applied Social Network Analysis theory to HCI \cite{Thimbleby}. Their work shows that an application's features can be interrelated using a social network. User's behavior represented as a social network graph is then able to unveil user's interactions and used features. Again, this application does not aim to provide automated user recommendations to interact with the application in an efficient way. It rather evaluates the work of the designer and gives hints which functions need more or less attention.

Bowen et al. utilize presentation interaction models (which are state transition systems) to describe graphical UIs and by modeling system manuals \cite{Bowen}. Their objective is to align and unveil inconsistencies between a graphical UI and according manuals. Though this work relates to user's training by providing correct training materials, it does not consider user's behavior logs and does not provide automated, step-wise training recommendations for users.

Dostal et al. developed a framework for the OpenOffice.org Suite which enables the logging of HCIs while interacting with OpenOffice \cite{Dostal:2011:RFP:1996461.1996511}. This tool enables many opportunities to analyze user's behavior in OpenOffice applications. However, it is limited to OpenOffice only and does not track HCI between applications and instances outside the OpenOffice.org Suite. Unfortunately, no research has been conducted on user's behavior logs observed by OpenOffice in order to recommend user's interactions.

Liu et al. discovered user's behavior from a software execution logs utilizing Process Mining techniques \cite{LiuUserBehavior}. However, the authors utilize the observed user's behaviors to detect patterns like commonly triggered user's operations in an application, specifically ProM, only.

In general, a significant amount of work has been done in the area of user's behavior analysis, but mainly in web analytics \cite{Kaushik:2007:WAH:1207733} and user activity tracking \cite{Atterer:2006:KUM:1135777.1135811}. Also, a lot of work has been done in web usage mining \cite{Chierichetti:2012:WUR:2187836.2187919, Mobasher:2000:APB:345124.345169}. Atterer et al. investigated the tracking of user's interaction using web technologies \cite{Atterer:2006:KUM:1135777.1135811}. The authors obtained meaningful statements by analyzing and refining the collected log files. Furthermore, there has been research conducted on utilization of Process Mining to detect abnormal behavior in social networks \cite{Sahlabadi}. A further approach to the analysis of farmers' interaction with decision support systems was shown by M\u{a}ru\c{s}ter et al. \cite{Maruster}.

There is also an area called \textit{Software Runtime Analysis} focusing on model generation \cite{Lorenzoli:2008:AGS:1368088.1368157} and mining \cite{Ammons:2002:MS:503272.503275, Gombotz}. However, these contributions do not deal with the user but with the runtime perspective.

All these areas utilize Process Mining to support the area of Software Development. To the best of our knowledge, no research has been conducted on supporting users to solve tasks in a more efficient way utilizing Process Mining techniques.

\section{Preliminaries}\label{sec:preliminaries}
In this section, we introduce the preliminaries and definitions that are required throughout this paper. We define event logs and Petri nets (PNs), provide an introduction to Process Mining and process discovery, and define HCI specific terms which are required to understand the proposed approach.

\subsection{Event Log}
The elementary component of \textit{event logs} are \textit{events}. An \textit{event} can be any real-life action consisting of a \textit{name} and an associated \textit{timestamp} \cite{VDAalst}. These two attributes are required whereas further attributes are optional. The term \textit{attribute} is often used as a synonym for \textit{resource} in Process Mining literature. Generally speaking, we define $\mathcal{E}$ representing the event space of all possible events. Moreover, $\mathcal{A}$ is the set of all possible attributes. Consequently, for $e \in \mathcal{E}$ and $a \in \mathcal{A}: \tau_a(e)$ is the value of the attribute $a$ \cite{definitions, VDAalst}. We define $a_{name}$ and $a_{time}$ as the attribute of an event's name and timestamp respectively.

A finite sequence of \textit{events} is defined as a \textit{trace} $g$ and is associated with an attribute \textit{name}. In any trace, each \textit{event} occurs only once and \textit{events} are ordered according to their \textit{timestamp} \cite{VDAalst, definitions}. The term \textit{case} is used synonymously for a \textit{trace}. We can define $\mathcal{T}$ as the set of all possible traces. Therefore, $g\in\mathcal{T}$ and $a\in\mathcal{A}:\tau_a(g)$ is the value of the attribute $a$ for trace $g$. Moreover, we define $|e|_g$ as the number of events in a trace $g$ \cite{definitions}. Therefore, the order of events in a trace has to satisfy
\begin{eqnarray}
\forall_{1\leq i < j\leq |e|_g} \tau_{a_{time}}(e_i) \leq \tau_{a_{time}}(e_j).
\end{eqnarray}
Finally, an \textit{event log} $L$ is a collection of \textit{traces} $g_i \in\mathcal{T}$ for $0\leq i $ such that each \textit{trace} $g_i$ occurs only once \cite{definitions}. 

\subsection{Process Mining}
Process Mining has been successfully applied in different fields like Healthcare \cite{PMtoHealth, Darabi}, Insurance \cite{PMtoInsurance}, Auditing \cite{PMtoAuditing}, and also Software Development \cite{PMtoSW, LiuTwoLayeredFramework, PMinOSS} and consists of three disciplines: (i) process discovery, (ii) conformance checking, and (iii) enhancement of process models \cite{VDAalst, PMtoSW}. A process model can be discovered by considering a log file as input. The output of such a discovery algorithm is a process model, which is usually a PN or a Business Process Modeling Notation (BPMN), Event Driven Process Chains (EPCs), or Casual Net (CNs). We are focusing on PNs only.

Conformance Checking is used to evaluate if a discovered process model is a good representation of the system recorded by a log. Two commonly used quality measures are \textit{fitness} and \textit{precision} among others. The \textit{fitness} metric measures if a process model can replay all events in each of an event log's traces. The \textit{precision} metric measures how well a model generalizes the actual process by simulation. Each generated trace by the process model should be a real-world trace, i.e. should be found in the actual event log.
Further measures include the complexity of a model. A process model is considered better the less complex it is. 

Lastly, enhancement considers a discovered process model as well as the logs to improve or extend the model using additional information, e.g. to predict events or to provide recommendations.

The term \textit{Software Process Mining} deals with analyzing software development processes, their runtime behavior as well as the user-software interaction \cite{LiuTwoLayeredFramework}. Therefore, we are mainly interested in the enhancement discipline of Process Mining, specifically in analyzing the user-software interaction or HCI. 

\subsection{Petri Net}
A PN is a commonly used modeling technique for process models. It consists of a set of places, graphically represented as circles and transitions represented as rectangles. Transitions usually correspond to events. The unidirectional interconnection between places to transitions and transitions to places is visualized using arcs. As such, a PN is defined as
\begin{eqnarray}
PN = (\mathcal{P}, T, F, \pi)
\end{eqnarray}
in which $\mathcal{P}$ is a set of places, $T$ is a set of transitions and $F \subseteq (\mathcal{P} \times T) \cup (T \times \mathcal{P})$ is a set of directed arcs connecting places and transitions \cite{definitions2}. $\pi$ is a function which maps transitions to event names. It is defined such that \cite{definitions2}
\begin{eqnarray}
\forall_{e \in \mathcal{E}} \exists_{t \in T} \pi(t) = \tau_{a_{name}}(e).
\end{eqnarray}
In general, each place can hold a non-negative integer number of tokens, however, in this paper the number of tokens in each place is either $1$ or $0$. We can define the number of tokens in a place $p$ as a function $\sigma(p)$ where $p \in \mathcal{P}$. For all $p \in \mathcal{P}$, the corresponding array of $\sigma(p)$ represents the marking $m$ of the PN. Each marking $m$ such that $m \in M$ represents the state of the PN  \cite{VDAalst}. We call the initial marking $m_{0}$, whereas the final marking is represented as $m_{z}$. Therefore, the sum over the vector elements of $m_0$ and $m_z$ must be each greater than $0$.
Furthermore, a transition $d \in T$ can only be fired if Equation (4) is satisfied for $p_{n} \in P: \exists (p_n \times d) \in F$ and if $p_{n} \neq \emptyset$.
\begin{eqnarray}
\forall p_n: \sigma(p_{n}) \geq 1
\end{eqnarray}\label{eq:1}
By firing a transition, a token is removed from each input place to that transition, while for each outgoing place from the transition the number of tokens will be increased by $1$. An important property to maintain is \textit{soundness} which describes the absence of livelocks, deadlocks, and other anomalies. Generally speaking, a PN is considered to be \textit{sound} if and only if for each trace \cite{soundness}:
\begin{itemize}
  \item it is always possible to reach the final marking,
  \item there are no remaining tokens when the final marking is reached,
  \item and if it is possible to execute an arbitrary event starting from $m_0$ by following the appropriate route through the PN.
\end{itemize}

\begin{figure}
  \begin{center}
    \includegraphics[width=\textwidth]{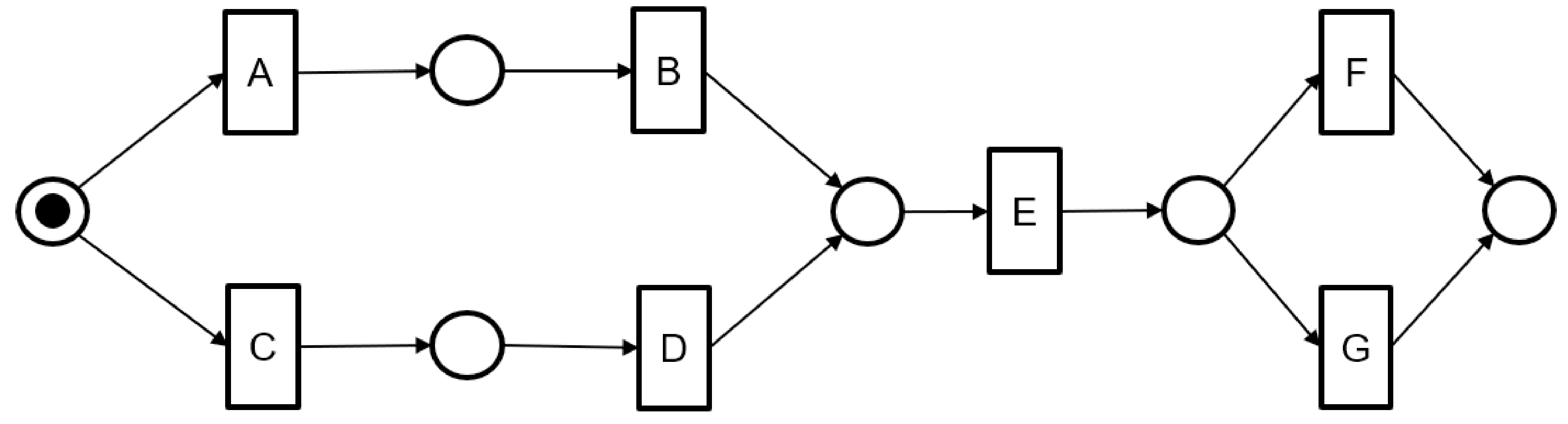}
  \end{center}
  \caption{Basic PN with seven transitions and six places. The first place holds a token.}
  \label{fig:basicpn}
\end{figure}

A basic PN can be seen in Figure \ref{fig:basicpn}. It holds a token in the first place, representing the initial marking $m_{0}$.

A special kind of transition is a so-called \textit{invisible transition}, graphically represented as a black rectangle. Invisible transitions can always be fired without corresponding events from a log as long as token requirements in incoming places are satisfied. Invisible transitions are used to model complex dependencies.

We call a PN that models user's behavior a \textit{behavioral PN}.
\subsection{Process Discovery Algorithm}
Throughout this paper, we use a state-of-the-art discovery algorithm called \textit{split miner} \cite{SplitMiner, splitminer2} which is a recent technique to discover PNs from event logs. The method has been developed by Augusto et al. with the objective to detect models with high fitness and precision, yet low complexity. The algorithm consists of five steps: a directly-follows dependency graph will be created and short loops will be discovered at first. Afterwards, the algorithm searches for concurrency and marks respective elements as such. In a third step, filtering will be applied such that each node is on a path from a single start to end node (maintaining soundness), the number of edges are minimal (minimizing complexity), and every path from start to end has the highest possible sum of frequencies (maximizing fitness). Afterwards, the algorithm adds split gateways in order to capture choice and concurrency. In a final step, joins will be discovered. 

The \textit{split miner} algorithm works well on a huge set of artificial and real-world event logs compared to existing methods \cite{PNBenchmark} and achieves significant performance improvements.

\subsection{Further Definitions}
In this paper, we are differentiating between \textit{high-level events} and \textit{low-level events}. We define \textit{low-level events} as events produced by software modules and libraries, such as Dynamic Link Libraries (DLLs). We call it \textit{low-level} since the event is recorded in its raw format. Therefore, \textit{low-level events} are usually hard to understand and to debug. In contrast, \textit{high-level events} are easier to understand, and follow the format of event logs which can be utilized for Process Mining. Thus, \textit{high-level events} consist of a \textit{name} as well as a \textit{timestamp} and optionally further attributes. \textit{Low-level events} can be translated to \textit{high-level events}.

Furthermore, we differentiate between \textit{controllable} and \textit{uncontrollable} events. \textit{Controllable events} are events which a user physically performs, such as keystrokes, mouse movements, or clicks. In contrast, we define \textit{uncontrollable events} as events which result from a sequence of \textit{controllable events} and underlying software logic. Examples are the visual appearance of an application window or closing of an application's instance. Both of the examples can be triggered by different sequences of \textit{controllable events}.

Finally, we are differentiating between \textit{Intra Task} and \textit{Inter Task} behavior. We are defining an Intra Task $t_{intra}$ as a sequence of controllable events which end with a single uncontrollable event. We denote controllable events as $c_i$, where $i$ is the $i$th controllable event in a sequence, and uncontrollable ones as $u$. Thus, $t_{intra} = \{c_1,...,c_i,u\}$ where $i > 0$. An example for an Intra Task could be changing the directory from a directory \textit{a} to its subfolder \textit{b}.
An Inter Task, denoted as $t_{inter}$, instead deals with the sequence of Intra Tasks, as such only with the uncontrollable events. An example could be the task to open a Notepad instance. This can be achieved by the sequence of the Intra Tasks \textit{open Windows menu}, \textit{search for Notepad.exe}, and \textit{open application}.

\section{Approach}\label{sec:approach}
The approach consists of two steps. First, we consider the acquisition of data since no suitable datasets are publicly available. The data acquisition section describes how we derive and develop an HCI user's behavior simulator for a given task from real participant observations. In the second part of this section, we propose the actual recommendation engine which is based on the output of the HCI user's behavior simulation. 

\subsection{Data Acquisition}\label{sec:dataacquisition}
\begin{figure}[h]
\begin{center}
  \includegraphics[width=210pt]{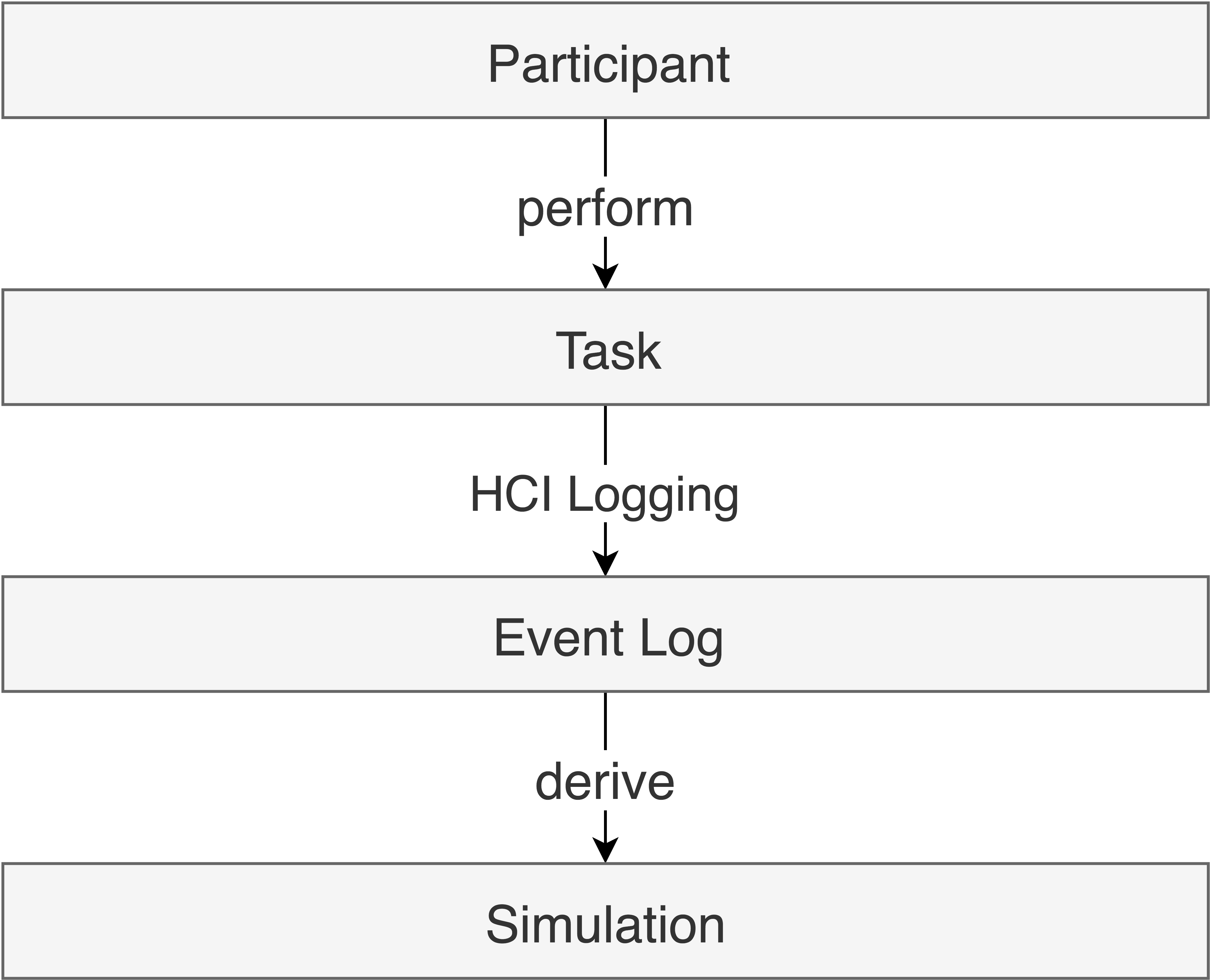}
  \caption{Flow diagram of the data acquisition process.}
  \label{fig:simulation}
\end{center}
\end{figure}

We are presenting the applicability of the proposed method based on a simple yet realistic real-world example in a classic desktop environment. We asked five real users/participants to solve a well-defined task while tracking their HCIs as low-level events. The resulting logs were used to create a simulation in order to create a larger amount of traces in a scalable way. The simulated traces were used to validate our approach. Figure \ref{fig:simulation} visualizes this procedure in a flow diagram.

Here, we describe the actual \textit{Task}, the \textit{HCI Logging}, and the final derived \textit{Simulation} in detail.

\subsubsection{Task}\label{sec:acquisitiondescription}
The participants are asked to summarize a company's performance data consisting of two files located in the \textit{documents/company data} folder for 15 times. Each of those two files consists of information about revenues and expenses of a single product. In order to solve the problem, the user has to create a simple text file called \textit{summary.txt} in the folder \textit{documents/company data/summaries} and has to fill in the summarized information. The applications a user can use are the \textit{Windows Calculator}, \textit{Windows Explorer}, \textit{Notepad} as well as the standard desktop. The user is provided a mouse, keyboard, screen, and a computer with a \textit{Windows 10} operating system.

\subsubsection{HCI Logging}\label{sec:logging}
To record the HCIs, we develop a Java application which tracks all inputs as well as visually shown on-screen events. The application tracks mouse movements, mouse clicks as well as keystrokes and leverages Window's DLLs to obtain information about opened, closed, maximized, and minimized applications. Additionally, our application tracks the hierarchy of visually on-screen shown applications and current folder locations of the \textit{Windows Explorer}. Table \ref{table-log-events} shows the details of the tracked events. Each event is associated with a corresponding timestamp. Moreover, in order to protect the privacy of participants, we only track the keys listed in Table \ref{table-keycodes}. All other keystrokes are expressed as a \textit{TEXT KEY} or \textit{NUM KEY}, depending if the text key was an alphabetical character or numerical/mathematical operation key. For simplicity reasons, we have split the screen into 4x4 rectangles such that mouse movements and clicks are not recorded pixel-wise but within the defined regions. The logging application is used to obtain HCI traces of the participants while performing the above described task.

The source code of the logging application is available in our Github repository \footnote{The URL will be made available upon acceptance.
}.

\begin{table}[h]
\centering
\resizebox{\textwidth}{!}{%
\begin{tabular}{lll}
\hline
\textbf{Event Key} & \textbf{Parameter}                                                                                                         & \textbf{Description}           \\ \hline
A1                 & \begin{tabular}[c]{@{}l@{}}{[}PID, ProcessName, WindowTitle, width, height, top\_left\_x, top\_left\_y{]}\end{tabular} & Application opened             \\
A2                 & {[}PID, ProcessName, WindowTitle{]}                                                                                        & Application closed             \\
A3                 & {[}PID, ProcessName{]}                                                                                                     & Application maximized          \\ 
A4                 & {[}PID, ProcessName{]}                                                                                                     & Application minimized          \\
A5                 & {[}PID, ProcessName, WindowTitle{]}                                                                                        & Window title changed           \\
A6                 & {[}PID, ProcessName, width, height, top\_left\_x, top\_left\_y{]}                                                          & Window position changed        \\ 
A7                 & {[}PID\#ProcessName;...;PID\#ProcessName{]}                                                                                 & Window hierarchy order changed \\
A8                 & {[}PID, OldPath, NewPath{]}                                                                                                & Explorer path changed          \\
K1                 & {[}KeyCode{]}                                                                                                              & Key pressed                    \\
K2                 & {[}KeyCode{]}                                                                                                              & Key released                   \\
K3                 & {[}KeyCode, x, y{]}                                                                                                        & Mouse click pressed            \\
K4                 & {[}KeyCode, x, y{]}                                                                                                        & Mouse click released           \\
K5                 & {[}-1 | 1{]}                                                                                                               & Mouse wheel                    \\
M                  & {[}x, y{]}                                                                                                                 & Mouse position    \\ \hline            
\end{tabular}%
}
\caption{All low-level events of the developed logging application. Each event consists of an event key as well as a set of parameters where PID is the application's Process Identifier used by the operating system.}
\label{table-log-events}
\end{table}

\begin{table}[h]
\begin{center}
\begin{tabular}{ccc}
\hline
\textbf{Type} & \textbf{Key Code}                                                 & \textbf{Description}           \\ \hline
Keyboard      & 29                                                                & Left ctrl                      \\
Keyboard      & 3613                                                              & Right ctrl                     \\
Keyboard      & 3675                                                              & Left meta                      \\ 
Keyboard      & 28                                                                & Enter                      \\
Keyboard      & 3665                                                              & Page down                     \\
Keyboard      & 3657                                                              & Page up                      \\

Keyboard      & 1                                                                & ESC                      \\
Keyboard      & 56                                                              & Left alt                    \\
Keyboard      & 3640                                                              & Right alt                      \\ 

Mouse      & 1                                                                & Left click                      \\
Mouse      & 2                                                              & Right click                    \\
Keyboard      & 42                                                              & Shift                      \\ 

Keyboard      & 54                                                                & Right shift                      \\
Keyboard      & 57419                                                              & Left arrow                     \\
Keyboard      & 57416                                                              & Top arrow                      \\ 

Keyboard      & 57421                                                                & Right arrow                      \\
Keyboard      & 57424                                                              & Down arrow                     \\
Keyboard      & 59-68                                                              & F1-F10                      \\ 
Keyboard      & 87-88                                                                & F11-F12                       \\
Keyboard      & 15                                                              & Tab                    \\
Keyboard      & 14                                                              & Backspace                     \\ 
\hline
\end{tabular}
\caption{Overview about the recorded keys and key codes.}
\label{table-keycodes}
\end{center}
\end{table}

\subsubsection{Simulation}\label{sec:dataacquisitionsimulation}
For Process Mining purposes, we need multiple traces per user in order to discover a process model. A small amount of traces would not provide sufficient behavioral insight. Therefore, we develop a logical simulation software with control parameters to create low-level event traces at scale. The development of this logical simulation is based on observations of the participant's event logs in order to create a larger set of traces. These large set of traces can be used to demonstrate the applicability of the proposed approach. The source code of the simulation is available in our Github repository, too \footnote{The URL will be made available upon acceptance.
}. 

The obtained low-level event logs of each real participant encompasses between $15,000$ and $30,000$ events over a time period of $18$ to $28$ minutes for the $15$ traces. We observe that each participant had an individual level of reactivity, which resulted in different speeds of mouse movements and frequencies of keystrokes. Therefore, one of the simulation parameter is \textit{Reactivity}. Similarly, each participant had a different mouse precision which resulted in the parameter \textit{MousePrecision}. Next, some participants tend to open new application instances instead of reopening already existing ones and they tend to close or minimize applications. These behaviors can be controlled using the parameters \textit{Minimize}, \textit{AppClosing} and \textit{AppOpenOrReopen}. Another observation is that some participants used the \textit{Windows Search Functionality} instead of browsing the \textit{Windows Explorer}. Therefore, we introduce a parameter \textit{Search}. Furthermore, some of the participants utilized hotkeys, whereas others solved the problem using the mouse only. This behavior is controlled with the parameter \textit{HotkeyUsage}. Finally, the task can be separated into subtasks such as \textit{creating summary file}, \textit{changing location to documents/data summaries}, \textit{calculating}, etc. These tasks can be repeated multiple times and in different orders. To model this behavior, we introduce the two parameters \textit{Repetition} and \textit{Sequential}. 
All parameters are expressed as the likelihood of the corresponding behavior with a numerical value between $0$ and $1$. The logical path simulation is developed in Python and creates traces based on the low-level events defined in Table \ref{table-log-events}.

Moreover, we develop a log translator which translates the low-level events of Table \ref{table-log-events} to high-level events. The reason for this is that even trivial switching between two applications produces multiple low-level events. Such a high-level event is always triggered by a user's interaction, which in turn is defined as a controllable event. User's interactions can be a mouse click or keystroke, or a sequence of clicks and keystrokes. Usually, a controllable event leads to an uncontrollable event which consists a sequence of the low-level events \textit{A1} to \textit{A7}. Switching between two applications leads for example to \textit{A3}, \textit{A6} and \textit{A7} simultaneously. Since such sequences would add extra complexity to PN models, we are translating them to high-level events. These high-level events are human-understandable, e.g. \textit{open explorer}, \textit{minimize notepad}, \textit{maximize calculator}, etc.. Later on, we will use high-level events to discover the behavioral PNs.

\subsection{Proposed Recommendation Engine}\label{sec:recommendationengine}
\begin{figure}[h]
\begin{center}
  \includegraphics[width=300pt]{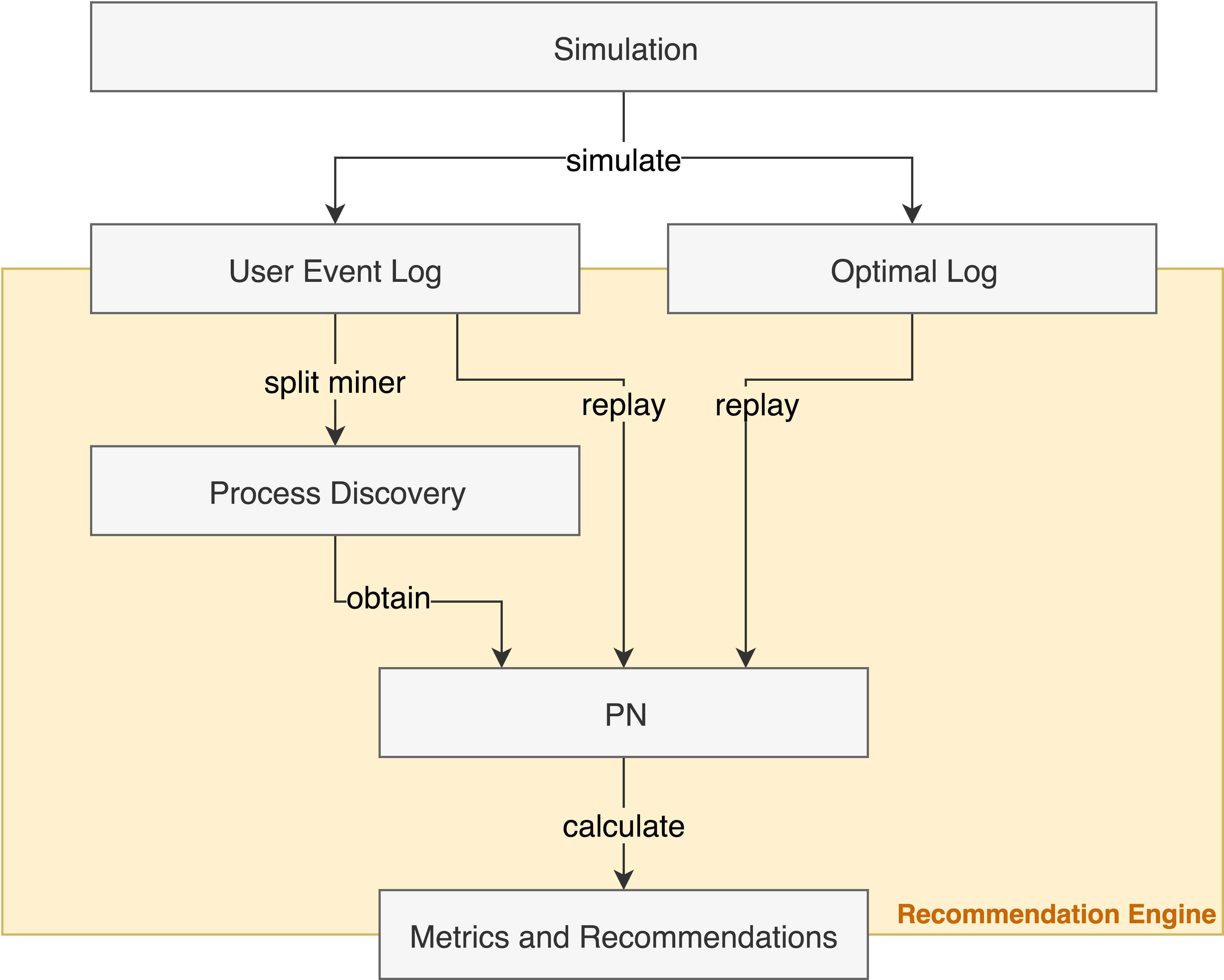}
  \caption{Flow diagram of the proposed approach.}
  \label{fig:rec_engine}
\end{center}
\end{figure}

In this section, we propose the actual recommendation engine. Figure \ref{fig:rec_engine} illustrates the procedure. We use the simulation described in the previous section to simulate a user's event log with specific characteristics, i.e. parameters. The user's event log has to consist of a statistically representative amount of traces of high-level events. The optimal log, however, consists only of one trace with high-level events. This trace is the optimal behavior we want to achieve. Therefore, the simulation parameters are set to best possible values. The user's event log is used to discover a behavioral PN using the \textit{split miner}. Finally, the optimal log and the user's event log are replayed against the obtained PN in order to calculate metrics and user recommendations. Especially the calculation of metrics and recommendations is discussed in this section.

In the first step, a behavioral PN will be discovered on the user's event log. This model summarizes the diverse set of traces and removes noise while being a statistical valid and solid representation of the user's behavior. Therefore, we utilize the high-level event traces from the HCI user simulation and leverage the earlier introduced \textit{split miner} discovery algorithm.

We replay the user's event log as well as the optimal log on the obtained behavioral PN. This is also called a \textit{conformance} check. Therefore, we utilize a replay function which considers only \textit{move on model} transitions. This means that the conformance metrics are calculated based on missing tokens in the PN only, such that we will never skip and penalize the actual events for fitness calculation purposes. The reason for this is that the user's behavior, and thus the traces, are the ground truth to be optimized rather than the PN model. If a user's behavior does not represent a specific optimal behavior, it should only move tokens within the model. Moreover, as long as the current state of the model is in a non-concurrency situation, there will be only one token in the system. If a deviation between a trace and the model exists, we are calculating the minimal distance between the current token position and the corresponding event observed in the log. The higher the distance, the higher the cost associated with the deviation. The cost for moving one token for one transition is $1$.

The fitness of the user's behavior event logs is measured as a baseline. For comparison purposes, the fitness of the optimal trace is measured, too. Since the \textit{split miner} discovery algorithm removes noisy behavior by filtering, we are not assuming a perfect fitness value. Instead we expect a value between 60\% and 90\%. The optimal trace fitness value will be lower assuming that the user's behavior is not optimal. The difference between the user's behavior and optimal fitness values provides a good indication of how close a user is to the optimal task solution.

Furthermore, the time span between transition enabling, i.e. the time Equation (4) is satisfied, and firing are measured. We also measure the frequency between transition enabling and actual transition firing. This information is important to obtain the reactivity of a user as well as to obtain his frequent behavior, in particular to differentiate between common and accidental behavior. The reactivity can be measured as the average firing time between two consecutive controllable events. In a very similar way, we can measure the mouse precision. The first and last mouse movement events in a consecutive sequence of mouse movements provide the start and end position of the cursor. Using this information, one can calculate the shortest path from start to end. The shortest path can be compared to the sequence replayed on the behavioral PN model. The paths are being compared to obtain a value corresponding to the user's mouse precision. Reactivity and mouse precision can be used to recommend specialized training. However, the reactivity value is an important metric to calculate the costs for further recommendations.

\begin{figure}[h]
\begin{center}
  \includegraphics[width=\textwidth]{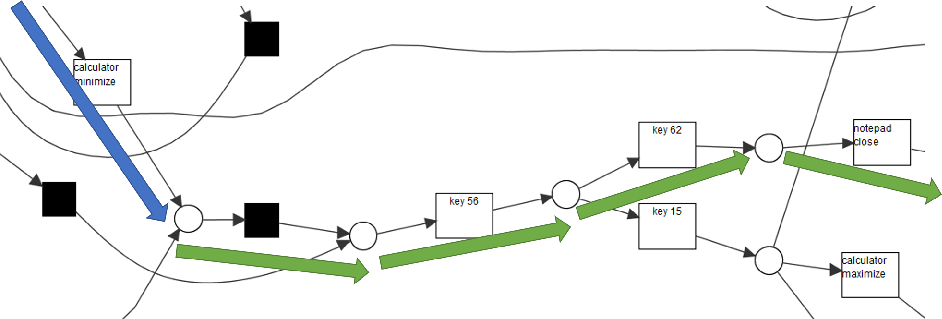}
  \caption{This figure shows an extract of a discovered PN from user's event logs and visualizes an Intra Task. The blue arrow indicates the uncontrollable event of the previous Intra Task. Green shows the complete flow of an Intra Task ending with the uncontrollable event \textit{notepad close}.}
  \label{fig:one}
\end{center}
\end{figure}

Moreover, we provide a user with Intra Task and Inter Task recommendations. 
An example for an Intra Task could be changing the directory from the \textit{documents} folder to its subfolder \textit{company data}. This Intra Task can be solved in a couple of ways using different sequences of controllable events. Figure \ref{fig:one} shows the path of a sample Intra Task in a PN.

An Intra Task recommendation consists of the final uncontrollable event as descriptor, a percental occurrence rate as the likelihood that the Intra Task occurs in a trace, the total number of occurrences averaged per trace, the average time saving based on the user's reactivity in milliseconds as well as the controllable event sequence a user should perform and the one the user should avoid, based on the optimal log. To obtain these recommendations, we first detect the Intra Tasks which deviate from the optimal log by replaying. We then calculate the cost which is the time a user could save by applying the optimal Intra Task behavior over the already applied user's behavior based on the earlier obtained user reactivity.

In comparison, an Inter Task is a sequence of Intra Tasks. Therefore, the shortest sequence of multiple $t_{intra}$ tasks is supposed to be the optimal solution to execute a specific task. As such, we analyze the Inter Task sequence obtained from the user's event log replay compared to the optimal log for repetitions. We calculate the cost for skipping Inter Tasks in order to argument how valuable a recommendation is to the user. The structure of Inter Task recommendations is very similar to Intra Task recommendations. It consists of a sequence of uncontrollable events which a user should not repetitively trigger. Moreover, it consists of a percental occurrence rate as the likelihood of occurrence per trace as well as an average time saving in milliseconds. 

The Java source code of the proposed approach has been published in our Github repository \footnote{The URL will be made available upon acceptance.
}.

\section{Results}\label{sec:results}
In order to evaluate our approach, we set up different user's behavior scenarios and create the corresponding event logs utilizing the developed simulation. Table \ref{table-userbehavior} provides an overview. The simulation is based on random sampling and every run is independent and identically distributed. When defining different user's behaviors, the likelihoods have been set to $0.05$ or $0.95$ accordingly. Therefore, we have chosen a sample size of $825$ traces to obtain a 95\% confidence interval with an interval size smaller than $1.5$ for each simulation parameter. Thus, we allow for enough random behavior, and obtain a statistical representative sample. 

\begin{table}[h]
\resizebox{\textwidth}{!}{%
\begin{tabular}{ccccl}
\hline
\textbf{User} & \textbf{Traces} & \textbf{Events} & \textbf{Time [s]} & \textbf{Description}           \\ \hline
Optimal      & 1   & 145.0 &  65.56                                                           & High precision and reactivity, no repetitions, hotkeys \\
User 1      & 825  & 250.31 & 209.58                                                          & Lower reactivity, low mouse precision  \\
User 2      & 825   & 189.00 & 75.58                                                          & Very low reactivity, no hotkey usage \\
User 3      & 825   & 159.04 &  81.00                                                        & Very low reactivity, repetitions\\
User 4      & 825  & 205.24 & 232.31                                                           & Low reactivity, repetitions, no hotkeys \\ 
User 5      & 825  & 446.74 & 191.03                                                           & High reactivity, low key precision, repetitions, no hotkeys \\
\hline
\end{tabular}
}
\caption{Overview of the modeled high-level user's event logs. The \textit{Events} and \textit{Time} column show the average number of events and average time of each trace.}
\label{table-userbehavior}
\end{table}
We run the above approach on the five different user's behaviors defined in Table \ref{table-userbehavior}. The measured reactivity of the optimal trace is 328ms. The optimal mouse precision is 1.0. Results are shown in Table \ref{table-results}. 

\begin{table}[h]
\centering
\resizebox{\textwidth}{!}{%
\begin{tabular}{cccccc}
\hline
\textbf{User's Behavior} & \textbf{User's Fitness} & \textbf{Optimal Fitness} & \textbf{User's Reactivity [ms]} & \textbf{User's Mouse Precision} \\ \hline
User 1                 & 0.804                  & 0.734                    & 674                      & 0.352                         \\
User 2                 & 0.884                  & 0.565                    & 901                      & 1.0                           \\
User 3                 & 0.888                  & 0.679                     & 922                      & 1.0                           \\
User 4                 & 0.918                 & 0.510                    & 731                      & 1.0                           \\
User 5                 & 0.874                  & 0.582                     & 337                      & 1.0   
\\ \hline
\end{tabular}%
}
\caption{Metrics obtained from the simulated user's behavior scenarios}
\label{table-results}
\end{table}

For \textit{User 1}, who is in general slower and less precise, a high user reactivity time of 674ms compared to the optimal 328ms is recorded. Also, the mouse precision of 0.328 is very low compared to 1.0. The user's fitness of 0.804 is as expected, but not perfect due to the already mentioned reasons in section \ref{sec:recommendationengine}. The fitness value of the optimal trace has a value of 0.734 and is quite high relatively to the assumptions we made. However, it is lower than the user fitness value due to imprecise mouse events.

\textit{User 2} does not use any hotkeys. Therefore, the user's behavior fitness value is at a baseline of 0.884. The optimal trace has a very low fitness value of 0.565 which makes sense as many of the optimal event sequences cannot be replayed on the user's behavior model. Reactivity is very low and mouse precision high, as simulated.

The reactivity and mouse precision of \textit{User 3} are as expected. However, this user's behavior is simulated such that he/she repeats Inter Tasks often. Therefore, the user's fitness as well as optimal fitness values are comparatively high with 0.888 and 0.679. This is due to the fact that both sequential strategies are very similar except for repetitions of Inter Tasks. Such repetitive behavior is not reflected in the fitness value.

The behavior of \textit{User 4} is a mix of everything. This user is slow, repeats a lot of Inter Tasks and does not consider hotkeys. Therefore, the user's fitness baseline is 0.918. The fitness value of the optimal trace is very low with 0.510 as expected. 

Finally, \textit{User 5} does not consider hotkeys and is simulated with a high likelihood of repetitive Inter Tasks. Similarly, the user's fitness is high with 0.874 while the optimal one is comparatively low. Both, reactivity and precision values are acceptable.

Intra Task and/or Inter Task Recommendations have been obtained for all user's behaviors except \textit{User 1}. As \textit{User 1} was simply slower and imprecise, but all other parameters were optimal, no further recommendations can be provided. As such, for example for \textit{User 2}, we observe the following Intra Task recommendation:
\begin{center}
\makebox{\textit{Task: explorer path to documents/summary}}\par
\makebox{\textit{Occurrence Rate: 0.460}}\par
\makebox{\textit{Total Occurrence Per Trace: 1.412}}\par
\makebox{\textit{Average Time Saving: 8109}}\par
\makebox{\textit{User should not do: [mouse click, key TEXT, key TEXT, key TEXT, key TEXT, key TEXT,}}\par
\makebox{\textit{key TEXT, key TEXT, key 28]}}\par
\makebox{\textit{User should do instead: [mouse doubleclick]}}
\end{center}

The user always typed the word \textit{summary} in the application window of the Explorer and pressed \textit{Enter} in order to move on into another directory. However, a simple double click would have been faster according to the user's reactivity.
Another example of an Intra Task recommendation for \textit{User 2} is the following:

\begin{center}
\makebox{\textit{Task: explorer maximize}}\par
\makebox{\textit{Occurrence Rate: 0.162}}\par
\makebox{\textit{Total Occurrence Per Trace: 2.725}}\par
\makebox{\textit{Average Time Saving: 3604}}\par
\makebox{\textit{User should not do: [mouse to 4,3, mouse to 3,3, mouse to 2,3, mouse to 1,3,}}\par
\makebox{\textit{mouse to 1,2, mouse click]}}\par
\makebox{\textit{User should do instead: [key 56, key 15]}}
\end{center}

Instead of maximizing the Explorer by moving the mouse and clicking, it would be faster for the user to use the combination of \textit{ALT} and \textit{TAB} on the keyboard. However, the occurrence rate as well as the average time saving are lower. Therefore, the first Intra Task has a higher priority than the second one due to the average time saving.

In this way, we obtained Intra Task recommendations for \textit{User 2}, \textit{User 4}, and \textit{User 5}. Especially, since those users have not utilized hotkeys. \textit{User 3}, however, has an optimal Intra Task behavior.

Similarly, for \textit{User 3}, the algorithm recommends the following two Inter Task recommendations:
\begin{center}
\makebox{\textit{Task: notepad close}}\par
\makebox{\textit{Occurrence Rate: 0.413}}\par
\makebox{\textit{Average Time Saving: 39579}}\par
\makebox{\textit{User should not repetitively do: notepad close}}
\end{center}
~\newline~
\begin{center}
\makebox{\textit{Task: calculator minimize}}\par
\makebox{\textit{Occurrence Rate: 0.192}}\par
\makebox{\textit{Average Time Saving: 4806}}\par
\makebox{\textit{User should not repetitively do: calculator minimize}}
\end{center}

Both Inter Tasks were observed in the user's event logs repetitively. We also obtain suitable Inter Task recommendations for \textit{User 4} and \textit{User 5} which would fix the behavior introduced in the user profiles in Table \ref{table-userbehavior}. By incorporating and considering each of the recommendations, a user will get closer to the optimal behavior, i.e. will be trained gradually.

\section{Discussion and Conclusion}\label{sec:conclusion}
We developed an approach to provide HCI user recommendations using Process Mining assuming that at least one optimal way of interaction is known. This approach deals with HCI traces in multi-instance and multi-application environments and provides users with useful recommendations. Applications can be found in any HCI user interaction improvement problem in which time is a crucial component and users are confronted with a large amount of repetitive tasks. We target supervisors or commissaries in organizations whose objective is to optimize HCI of employees in a cost-effective manner. We successfully applied our Process Mining approach on simulated user traces based on a realistic example which has been designed and run with real users in advance. Therefore, our approach can be used to detect users who do not interact optimally, i.e. their user logs deviate from optimal ones. These users can be trained in a step-wise manner by providing them with either a single or few recommendations at a time. 

A re-validation of their HCI behavior and the impact of providing recommendations could be investigated at a later time to monitor each user's progress. In conclusion, the advantages for organizations are twofold: firstly, users will not be overtaxed and employee's satisfaction is still ensured. Secondly, organizations themselves increase efficiency while decreasing cost.

Our approach has a few limitations. Process Mining requires large amount of traces. Therefore, organizations have to collect data over a few weeks or even months, depending on the frequency of repetitive user tasks. Although, the translation of low-level to high-level events is expensive and requires domain knowledge. Significant effort has to be spent to track essential events.

In the future, an investigation on the performance and limitations of the proposed approach should be performed based on a real-world scenario and real event logs. Moreover, our proposed approach requires a logging application which is able to track HCIs in multi-application environments and to translate low-level events to high-level events. Unfortunately, there are no standardized logging interfaces for HCI optimization available which could be incorporated with approaches like ours. Therefore, research studies on the design and profitability of such HCI logging interfaces should be performed.

\bibliographystyle{unsrt}  

\begin{thebibliography}{10}

\bibitem{TowardsModernInclusiveFactories}
Valeria Villani, Lorenzo Sabattini, Julia~N Czerniaki, Alexander Mertens,
  Birgit Vogel-Heuser, and Cesare Fantuzzi.
\newblock Towards modern inclusive factories: A methodology for the development
  of smart adaptive human-machine interfaces.
\newblock In {\em 2017 22nd IEEE International Conference on Emerging
  Technologies and Factory Automation (ETFA)}, pages 1--7. IEEE, 2017.

\bibitem{VDAalst}
Wil M.~P. van~der Aalst.
\newblock {\em Process Mining: Discovery, Conformance and Enhancement of
  Business Processes}.
\newblock Springer Publishing Company, Incorporated, 1st edition, 2011.

\bibitem{PMtoSW}
Vladimir~A. Rubin, Alexey~A. Mitsyuk, Irina~A. Lomazova, and Wil M.~P. van~der
  Aalst.
\newblock Process mining can be applied to software too!
\newblock In {\em Proceedings of the 8th ACM/IEEE International Symposium on
  Empirical Software Engineering and Measurement}, ESEM '14, pages 57:1--57:8,
  New York, NY, USA, 2014. ACM.

\bibitem{LiuTwoLayeredFramework}
Cong Liu, Jianpeng Zhang, Guangming Li, Shangce Gao, and Qingtian Zeng.
\newblock A two-layered framework for the discovery of software behavior: A
  case study.
\newblock {\em IEICE TRANSACTIONS on Information and Systems},
  101(8):2005--2014, 2018.

\bibitem{ERP}
Thomas~H Davenport.
\newblock {\em Mission critical: realizing the promise of enterprise systems}.
\newblock Harvard Business Press, 2000.

\bibitem{Park2010}
Sung Park, Arthur~D. Fisk, and Wendy~A. Rogers.
\newblock Human factors consideration for the design of collaborative machine
  assistants.
\newblock {\em Handbook of Ambient Intelligence and Smart Environments}, pages
  961--984, 2010.

\bibitem{Shneiderman:2009:DUI:1593001}
Ben Shneiderman, Catherine Plaisant, Maxine Cohen, and Steven Jacobs.
\newblock {\em Designing the User Interface: Strategies for Effective
  Human-Computer Interaction}.
\newblock Addison-Wesley Publishing Company, USA, 5th edition, 2009.

\bibitem{TaskBasedUIDesign}
Martijn Van~Welie.
\newblock Task-based user interface design.
\newblock {\em SIKS Dissertation Series}, 6, 2001.

\bibitem{CALISIR2004505}
Fethi Calisir and Ferah Calisir.
\newblock The relation of interface usability characteristics, perceived
  usefulness, and perceived ease of use to end-user satisfaction with
  enterprise resource planning (erp) systems.
\newblock {\em Computers in human behavior}, 20(4):505--515, 2004.

\bibitem{Dev:2017:IFU:3025171.3025184}
Himel Dev and Zhicheng Liu.
\newblock Identifying frequent user tasks from application logs.
\newblock In {\em Proceedings of the 22Nd International Conference on
  Intelligent User Interfaces}, IUI '17, pages 263--273, New York, NY, USA,
  2017. ACM.

\bibitem{Cronholm:2009:UUG:1738826.1738864}
Stefan Cronholm.
\newblock The usability of usability guidelines: A proposal for
  meta-guidelines.
\newblock In {\em Proceedings of the 21st Annual Conference of the Australian
  Computer-Human Interaction Special Interest Group: Design: Open 24/7}, OZCHI
  '09, pages 233--240, New York, NY, USA, 2009. ACM.

\bibitem{Nielsen:1994:UE:2821575}
Jakob Nielsen.
\newblock {\em Usability Engineering}.
\newblock Morgan Kaufmann Publishers Inc., San Francisco, CA, USA, 1993.

\bibitem{Jacko:2012:HIH:2378709}
Julie~A. Jacko.
\newblock {\em Human-Computer Interaction Handbook: Fundamentals, Evolving
  Technologies, and Emerging Applications}.
\newblock CRC Press, Inc., Boca Raton, FL, USA, 3rd edition, 2012.

\bibitem{AUGUR}
Melanie Hartmann, Daniel Schreiber, and Max M\"{u}hlh\"{a}user.
\newblock Augur: Providing context-aware interaction support.
\newblock In {\em Proceedings of the 1st ACM SIGCHI Symposium on Engineering
  Interactive Computing Systems}, EICS '09, pages 123--132, New York, NY, USA,
  2009. ACM.

\bibitem{Palanque}
Philippe Palanque, Eric Barboni, C{\'e}lia Martinie, David Navarre, and Marco
  Winckler.
\newblock A model-based approach for supporting engineering usability
  evaluation of interaction techniques.
\newblock In {\em Proceedings of the 3rd ACM SIGCHI Symposium on Engineering
  Interactive Computing Systems}, EICS '11, pages 21--30, New York, NY, USA,
  2011. ACM.

\bibitem{Thimbleby}
Harold Thimbleby and Patrick Oladimeji.
\newblock Social network analysis and interactive device design analysis.
\newblock In {\em Proceedings of the 1st ACM SIGCHI Symposium on Engineering
  Interactive Computing Systems}, EICS '09, pages 91--100, New York, NY, USA,
  2009. ACM.

\bibitem{Bowen}
Judy Bowen and Steve Reeves.
\newblock Modelling user manuals of modal medical devices and learning from the
  experience.
\newblock In {\em Proceedings of the 4th ACM SIGCHI Symposium on Engineering
  Interactive Computing Systems}, EICS '12, pages 121--130, New York, NY, USA,
  2012. ACM.

\bibitem{Dostal:2011:RFP:1996461.1996511}
Martin Dost\'{a}l and Zdenek Eichler.
\newblock A research framework for performing user studies and rapid
  prototyping of intelligent user interfaces under the openoffice.org suite.
\newblock In {\em Proceedings of the 3rd ACM SIGCHI Symposium on Engineering
  Interactive Computing Systems}, EICS '11, pages 153--156, New York, NY, USA,
  2011. ACM.

\bibitem{LiuUserBehavior}
Cong Liu, Shi Wang, Shangce Gao, Feng Zhang, and Jiujun Cheng.
\newblock User behavior discovery from low-level software execution log.
\newblock {\em IEEJ Transactions on Electrical and Electronic Engineering},
  13(11):1624--1632, 2018.

\bibitem{Kaushik:2007:WAH:1207733}
Avinash Kaushik.
\newblock {\em Web analytics: an hour a day}.
\newblock John Wiley \& Sons, 2007.

\bibitem{Atterer:2006:KUM:1135777.1135811}
Richard Atterer, Monika Wnuk, and Albrecht Schmidt.
\newblock Knowing the user's every move: User activity tracking for website
  usability evaluation and implicit interaction.
\newblock In {\em Proceedings of the 15th International Conference on World
  Wide Web}, WWW '06, pages 203--212, New York, NY, USA, 2006. ACM.

\bibitem{Chierichetti:2012:WUR:2187836.2187919}
Flavio Chierichetti, Ravi Kumar, Prabhakar Raghavan, and Tamas Sarlos.
\newblock Are web users really markovian?
\newblock In {\em Proceedings of the 21st International Conference on World
  Wide Web}, WWW '12, pages 609--618, New York, NY, USA, 2012. ACM.

\bibitem{Mobasher:2000:APB:345124.345169}
Bamshad Mobasher, Robert Cooley, and Jaideep Srivastava.
\newblock Automatic personalization based on web usage mining.
\newblock {\em Commun. ACM}, 43(8):142--151, August 2000.

\bibitem{Sahlabadi}
Mahdi Sahlabadi, {Ravie Chandren} Muniyandi, and Zarina Shukur.
\newblock Detecting abnormal behavior in social network websites by using a
  process mining technique.
\newblock {\em Journal of Computer Science}, 10(3):393--402, 2014.

\bibitem{Maruster}
Laura M\u{a}ru\c{s}ter, Niels Faber, Ren\'{e} J.~Jorna, and Rob Van~Haren.
\newblock A process mining approach to analyse user behaviour.
\newblock {\em WEBIST 2008 - 4th International Conference on Web Information
  Systems and Technologies, Proceedings}, 2:208--214, 01 2008.

\bibitem{Lorenzoli:2008:AGS:1368088.1368157}
Davide Lorenzoli, Leonardo Mariani, and Mauro Pezz\`{e}.
\newblock Automatic generation of software behavioral models.
\newblock In {\em Proceedings of the 30th International Conference on Software
  Engineering}, ICSE '08, pages 501--510, New York, NY, USA, 2008. ACM.

\bibitem{Ammons:2002:MS:503272.503275}
Glenn Ammons, Rastislav Bod\'{\i}k, and James~R. Larus.
\newblock Mining specifications.
\newblock In {\em Proceedings of the 29th ACM SIGPLAN-SIGACT Symposium on
  Principles of Programming Languages}, POPL '02, pages 4--16, New York, NY,
  USA, 2002. ACM.

\bibitem{Gombotz}
Robert Gombotz and Schahram Dustdar.
\newblock On web services workflow mining.
\newblock In Christoph~J. Bussler and Armin Haller, editors, {\em Business
  Process Management Workshops}, pages 216--228, Berlin, Heidelberg, 2006.
  Springer Berlin Heidelberg.

\bibitem{definitions}
Qinlong Guo, Lijie Wen, Jianmin Wang, Zhiqiang Yan, and Philip~S. Yu.
\newblock Mining invisible tasks in non-free-choice constructs.
\newblock In Hamid~Reza Motahari-Nezhad, Jan Recker, and Matthias Weidlich,
  editors, {\em Business Process Management}, pages 109--125, Cham, 2015.
  Springer International Publishing.

\bibitem{PMtoHealth}
Ronny~S. Mans, Helen Schonenberg, Minseok Song, Wil M.~P. van~der Aalst, and
  Piet J.~M. Bakker.
\newblock Application of process mining in healthcare -- a case study in a
  dutch hospital.
\newblock In Ana Fred, Joaquim Filipe, and Hugo Gamboa, editors, {\em
  Biomedical Engineering Systems and Technologies}, pages 425--438, Berlin,
  Heidelberg, 2009. Springer Berlin Heidelberg.

\bibitem{Darabi}
Houshang Darabi, William~L. Galanter, Janet~Y. Lin, Ugo Buy, and Rupa Sampath.
\newblock Modeling and integration of hospital information systems with petri
  nets.
\newblock In {\em 2009 IEEE/INFORMS International Conference on Service
  Operations, Logistics and Informatics}, pages 190--195, July 2009.

\bibitem{PMtoInsurance}
Suriadi Suriadi, Moe~T. Wynn, Chun Ouyang, Arthur~H.M. ter Hofstede, and Nienke
  van Dijk.
\newblock Understanding process behaviours in a large insurance company in
  australia : a case study.
\newblock In {\em 25th International Conference on Advanced Information Systems
  Engineering, CAiSE 2013}, pages 449--464, Valencia, Spain, 2013. Springer.

\bibitem{PMtoAuditing}
Mieke Jans, Michael Alles, and Miklos Vasarhelyi.
\newblock The case for process mining in auditing: Sources of value added and
  areas of application.
\newblock {\em International Journal of Accounting Information Systems},
  14(1):1 -- 20, 2013.

\bibitem{PMinOSS}
Elia Kouzari, Lazaros Sotiriadis, and Ioannis Stamelos.
\newblock Process mining for process conformance checking in an oss project: An
  empirical research.
\newblock In Ioannis Stamelos, Jesus~M. Gonzalez-Baraho{\~{n}}a, Iraklis
  Varlamis, and Dimosthenis Anagnostopoulos, editors, {\em Open Source Systems:
  Enterprise Software and Solutions}, pages 79--89, Cham, 2018. Springer
  International Publishing.

\bibitem{definitions2}
A.~Adriansyah, B.~F. van Dongen, and W.~M.~P. van~der Aalst.
\newblock Conformance checking using cost-based fitness analysis.
\newblock In {\em 2011 IEEE 15th International Enterprise Distributed Object
  Computing Conference}, pages 55--64, Aug 2011.

\bibitem{soundness}
W.~M.~P. van~der Aalst, K.~M. van Hee, A.~H.~M. ter Hofstede, N.~Sidorova,
  H.~M.~W. Verbeek, M.~Voorhoeve, and M.~T. Wynn.
\newblock Soundness of workflow nets: classification, decidability, and
  analysis.
\newblock {\em Formal Aspects of Computing}, 23(3):333--363, May 2011.

\bibitem{SplitMiner}
Adriano Augusto, Raffaele Conforti, Marlon Dumas, and Marcello La~Rosa.
\newblock Split miner: Discovering accurate and simple business process models
  from event logs.
\newblock In {\em 2017 IEEE International Conference on Data Mining (ICDM)},
  pages 1--10, Nov 2017.

\bibitem{splitminer2}
Adriano Augusto, Raffaele Conforti, Marlon Dumas, Marcello La~Rosa, and Artem
  Polyvyanyy.
\newblock Split miner: automated discovery of accurate and simple business
  process models from event logs.
\newblock {\em Knowledge and Information Systems}, pages 1--34, 2018.

\bibitem{PNBenchmark}
Adriano Augusto, Raffaele Conforti, Marlon Dumas, Marcello La~Rosa, Fabrizio~M
  Maggi, Andrea Marrella, Massimo Mecella, and Allar Soo.
\newblock Automated discovery of process models from event logs: Review and
  benchmark.
\newblock {\em IEEE Transactions on Knowledge and Data Engineering}, 2018.

\end{thebibliography}

\end{document}